\documentclass{article}
\usepackage{spconf}
\usepackage{amsmath,amsthm,amsfonts,amssymb}
\usepackage{mathtools}
\usepackage[english]{babel}
\usepackage{float}
\usepackage[pdftex]{graphicx}
\usepackage{color}
\usepackage{multirow}
\usepackage{graphicx}
\usepackage{IEEEtrantools}
\usepackage{bm}
\usepackage{balance}
\usepackage[caption=false,font=footnotesize]{subfig}
\usepackage{url}
\usepackage{cite}
\usepackage{comment}
\usepackage[switch]{lineno}
\usepackage[named]{algo}
\usepackage[noend]{algpseudocode}
\usepackage{algorithmicx}
\usepackage{algorithm}
\usepackage{algpseudocode}
\algnewcommand\algorithmicinput{\textbf{INPUTS:}}
\algnewcommand\INPUTS{\item[{\textbf{STAGE 1: COMPUTING INITIAL CONDITIONS}}]}
\algnewcommand\STAGEA{\item[{\textbf{STAGE 2: LINEAR PROGRAM}}]}
\algnewcommand\STAGEB{\item[{\textbf{STAGE 3: CONJUGATE GRADIENT ITERATIONS}}]}
\usepackage{array,multirow}
\usepackage{nicefrac}
\usepackage{mathtools}
\usepackage{multicol}
\usepackage{subfig} % make it possible to include more than one captioned figure/table in a single float
\usepackage[font=small,skip=0pt]{caption}

\ninept

\DeclareMathOperator{\Tr}{Tr}
\DeclareMathOperator{\vectz}{vec}

\title{Phase Retrieval for Rydberg Quantum Arrays}
\name{Peter Vouras$^{\dag}$, Kumar Vijay Mishra$^{\ddag}$, Alexandra Artusio-Glimpse$^{*}$%~\IEEEmembership{Member,~IEEE}
\thanks{K. V. M. acknowledges support from the National Academies of Sciences, Engineering, and Medicine via Army Research Laboratory Harry Diamond Distinguished Fellowship.}
\address{$^{\dag}$United States Department of Defense, Washington, DC, 20375 USA\\
$^{\ddag}$United States DEVCOM Army Research Laboratory, Adelphi, MD 20783 USA\\
$^{*}$National Institute of Standards and Technology, Boulder, CO, 80303 USA
}
}

\begin{document}
% Reduce spacing above and below equations
\setlength{\abovedisplayskip}{3pt}
\setlength{\belowdisplayskip}{3pt}

\maketitle

\begin{abstract}
In this paper we derive a novel phase retrieval algorithm for use in phased array or synthetic aperture applications where only measurements of electric field intensity are possible at each spatial sample.  Such array configurations exist if a Rydberg atom probe is used in place of an antenna.  We present outcomes of numerical experiments showing the effectiveness of the proposed algorithm.
 %  Rydberg atoms and sensing ....
 
%  This process may be viewed as a signal reconstruction with some variation of phase retrieval methods. 
 
%  We provide a trust-region algorithm that minimizes a smoothed non-convex least-squares objective function to iteratively recover the underlying signal-of-interest for either time- or band-limited support. 
 
%  Our theoretical analysis shows that unique signal reconstruction is possible using signal samples no more than .....

\end{abstract}


\section{Introduction}
Phase retrieval is the process of recovering a signal from the magnitude of its Fourier transform. This problem has been studied in a number of applications, such as optics \cite{walther1963question}, X-ray crystallography \cite{harrison1993phase,millane1990phase}, speech recognition \cite{juang1993fundamentals}, blind channel estimation \cite{baykal2004blind}, and astronomy \cite{fienup1987phase}.  The interest in phase retrieval is largely due to the inability of a sensing device to measure the phase of a received signal.  For one-dimensional (1D) signals, the problem is ill-posed, meaning more than one signal with different phase can map to the same magnitude; the only exceptions being a minimum-phase signal \cite{Huang2015} or a sparse signal with structured support \cite{ranieri2013phase}.  Substantial work has been done and is still ongoing to overcome the ill-posedness of the problem, and the literature is too large to summarize here (e.g. see \cite{shechtman2014phase,jaganathan2015phase} for contemporary surveys, and references therein). Overall, two major approaches have emerged: the first harnesses prior knowledge of the signal structure, such as sparse support \cite{shechtman2011sparsity,jaganathan2013sparse,shechtman2014gespar}, while the other exploits technology to make additional measurements of the magnitude via, for example, masks \cite{candes2014phase} and short-time Fourier transform (STFT) \cite{eldar2015sparse,jaganathan2015stft,bendory2016non}.

Recent work at the National Institute of Standards and Technology (NIST) and many other institutions has demonstrated the feasibility of using Rydberg atom probes to provide traceable measurements of electric field intensity~\cite{Artusio2022}. These quantum probes have many unique features that set them apart from traditional antennas including exactly defined response functions based on quantum mechanics allowing for direct field strength traceability to the International System of Units (SI)~\cite{Gordon2010,Sedlacek2012,Holloway2014_Dec,Gordon2014,Anderson2021}, a highly transparent probe head to RF fields made only of dielectric materials~\cite{Simons2018_July,Anderson2022_patent}, direct down-conversion of the RF field by the atoms reducing the need for back-end electronics~\cite{Holloway2019_May}, and intrinsic ultra-wideband tunability from kilohertz frequencies to terahertz frequencies using a single probe and a tunable laser~\cite{Holloway2014_Dec,Meyer2021}. These unique features have attracted a large number of RF scientists, engineers, and developers to these Rydberg atom probes; however, one limitation of these probes is in the detection of phase. Typically, in order to resolve the phase of the incident RF field, a second RF field acting as a local-oscillator (LO) is radiated onto the Rydberg atom probe~\cite{Simons2019,Jing2020}. The radiated LO has implications for the transparency of the probe because it requires an additional antenna as well as impacting the usefulness of these probes in stealth applications. One alternative phase sensing approach has been put forward~\cite{Anderson2022} wherein an optical LO is used rather than a radiated RF LO; however, this scheme is limited to sensing only a single quadrant of phase space because it is essentially still an intensity detector. Thus, processing schemes that recover phase information from a purely intensity set of measurements are of interest for applications that use the Rydberg atom probes in a synthetic aperture measurement (e.g. \cite{Simons2018_July,Cardman2020}).

The algorithm described in this paper is intended to retrieve an estimate of signal phase from intensity-only measurements such that the Rydberg atom probe can be used to construct a synthetic aperture.  Synthetic apertures increase the effective area, and thereby angular resolution, of an imaging system beyond the physical limits of a hardware antenna.  A synthetic aperture is constructed by using a mechanical positioner to move a measurement probe or antenna through space.  The probe measures at discrete locations the RF fields propagating across the aperture and if the spatial RF samples maintain phase coherence then they can be combined in post-processing to create high-resolution snapshots of the ambient scattering. %  Primary examples of this approach include synthetic aperture radar (SAR), synthetic aperture channel sounding, and synthetic aperture radiometry. 

\section{Measurement Model}
\label{sec:system}
\begin{figure}
 \begin{minipage}[b]{1.0\linewidth}
  \centering
  \includegraphics[width=6cm]{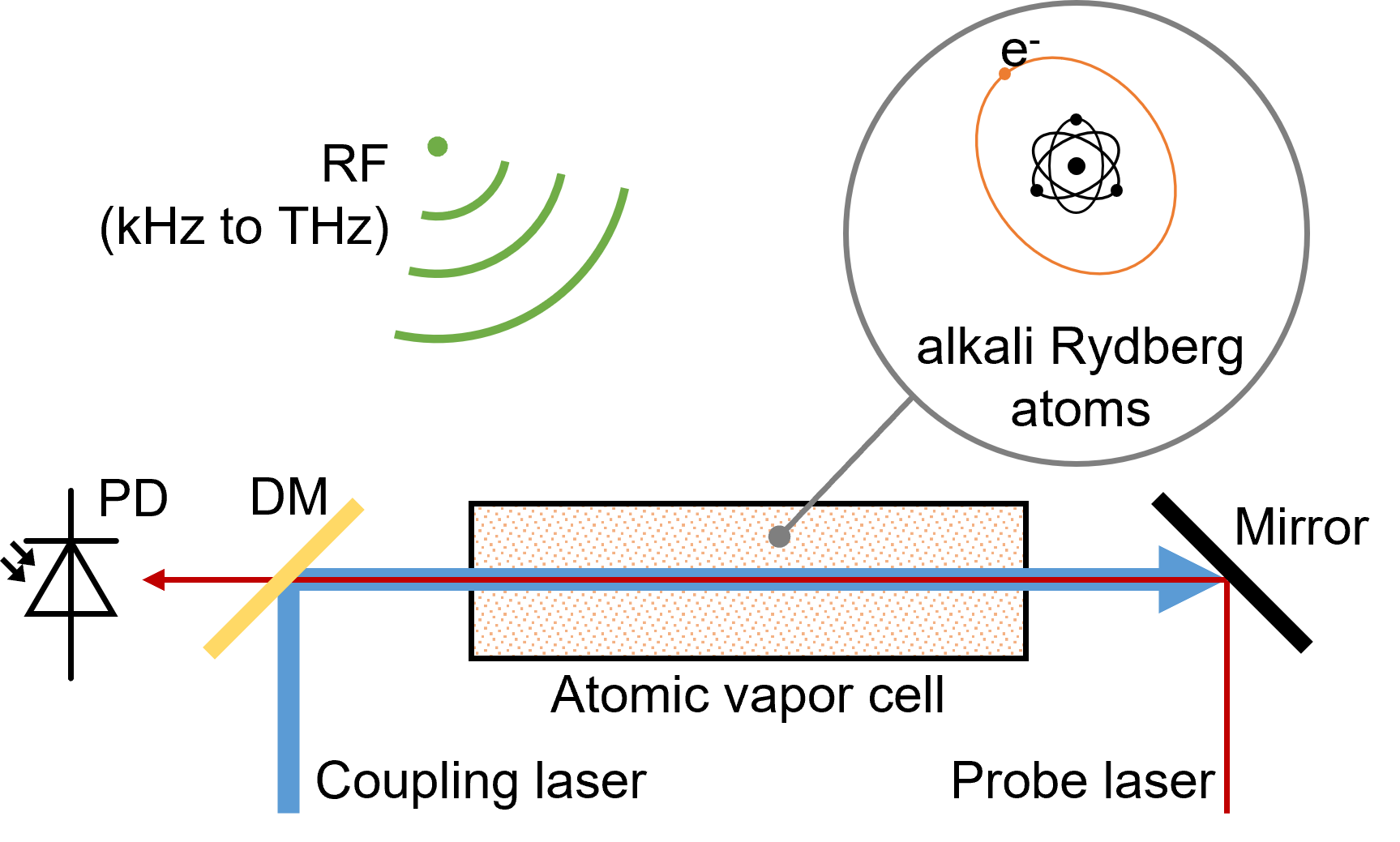}
  \vspace{0.1cm}
  \caption{Diagram of the Rydberg atom electric field measurement setup. PD is a photodetector and DM is a dichroic mirror. This same set of hardware can be used to detect RF electric fields from kHz to THz.}%\medskip
  \label{fig:rydbergSetup}
 \end{minipage}
\end{figure}

\begin{figure}
 \begin{minipage}[b]{1.0\linewidth}
  \centering
  \includegraphics[width=8.5cm]{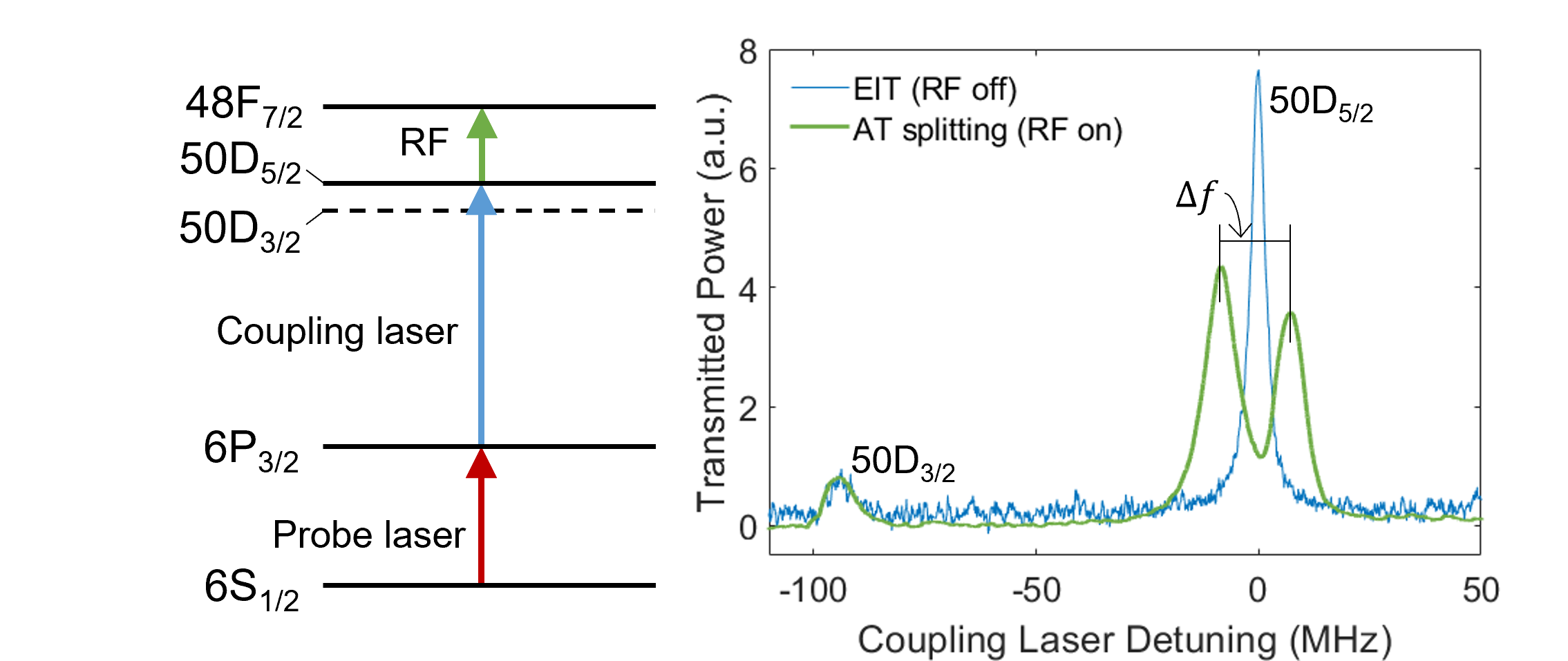}
  \vspace{0.1cm}
  \caption{Example Rydberg measurement when a 26.105~GHz RF field radiates $^{85}$Rb atoms excited to the $50D_{5/2}$ Rydberg state by the probe laser (resonant with the $6S_{1/2}$ and $6P_{3/2}$ states) and the coupling laser (resonant with the $6P_{3/2}$ and $50D_{5/2}$ states). When the coupling laser frequency is detuned, the EIT line of the probe transmittance appears (blue line). When the coupling laser is far detuned in the negative frequency direction, another weaker EIT line appears from the $50D_{3/2}$ state, which is used to calibrate the frequency axis of the spectrum. Once the resonant RF field is turned on, the main EIT line splits (green line) and the frequency difference, $\Delta_f$, is directly proportional to the field strength of the RF and the dipole moment of the $50D_{5/2}$ to $48F_{7/2}$ transition, Eq.~\ref{eq:rydbergEquation}.}%\medskip
  \label{fig:eitAT}
 \end{minipage}
\end{figure}

The Rydberg atom probe measurement utilizes sensitive Rydberg states (high energy states where the electron of an alkali atom is far from and weakly coupled to the nucleus) to detect RF electric fields. These Rydberg states are reached using a set of lasers. A near-infrared probe laser couples to the ground state of the atoms and is detected with a photodetector after passing through the atomic vapor cell, as depicted in Fig.~\ref{fig:rydbergSetup}. Because the probe laser is resonant with ground states of the atoms, this laser is preferentially absorbed and minimal light reaches the photodetector. Then, a counter-propagating visible coupling laser excites the atoms to a Rydberg state where the principle quantum number is large and the energy separation between adjacent Rydberg states is small (e.g. on the order of RF frequencies). The combination of these two lasers produces an effect known as electromagnetically induced transparency (EIT), wherein the atoms become transparent to the probe laser over a narrow frequency range and the light reaching the detector increases, see Fig.~\ref{fig:eitAT}. When the atoms are then radiated by an RF field that is resonant with a second Rydberg state, the narrow EIT line splits known as Autler-Townes (AT) splitting. The frequency difference of this AT line splitting, $\Delta_f$, is proportional to the absolute value of the electric field of the RF, $|E_{RF}|$:
\begin{equation}\label{eq:rydbergEquation}
    \Delta_f = \frac{\wp}{h}|E_{RF}|,
\end{equation}
where $\wp$ is the dipole moment of the Rydberg transition, which is accurately calculated \cite{Simons2016}, and $h=6.62607015\times10^{-34}~\mathrm{J~Hz^{-1}}$ is Planck's constant, a fundamental unit in the new SI. Once the on-resonance (detuning of 0~MHz in Fig.~\ref{fig:eitAT}) change in probe laser transmittance with RF field strength is calibrated, both laser frequencies can be locked and fast sensing of the RF field can be achieved. Our paper considers this Rydberg atom RF field measurement system when the sensor, the vapor cell and lasers packed with fiber-coupling~\cite{Simons2018_July}, is scanned over a spatial plane defining a synthetic aperture and measurements of the RF electric field intensity are captured at each location. Due to the absolute value in Eq.~\ref{eq:rydbergEquation}, no phase information of the RF field is captured and must be retrieved using the processing described below.

%\textcolor{red}{We cannot directly write this equation. There are many assumptions and specifics that need to be specified before this equation. See, e.g., how I write the text before eq. (5) of my paper: \url{https://arxiv.org/pdf/1802.09736.pdf} }
A common approximation, also used here, is that signal sources in the scene are far enough away from the observation plane of the aperture that the impinging phase front is nearly planar.  An additional assumption is that the signal is narrowband, e.g. a sinusoidal tone.  With these conditions satisfied, a linear signal model describes the signals propagating across an array of ${N}$ elements as a sum of plane waves, or ${\mathbf{A}\mathbf{s} = \mathbf{b}}$.

The columns of ${\mathbf{A}}$ are steering vectors sampled on a discrete search grid of ${K}$ angles ${(u_{i},v_{i})}$ defined in sine space coordinates,
\begin{equation}
\label{E:eqn02}
\mathbf{A} = \left[ \begin{array}{ccc} \mathbf{a}(u_0,v_0) & \ldots & \mathbf{a}(u_{K-1},v_{K-1}) \end{array} \right]
\end{equation}
with
\begin{equation}
    \mathbf{a}(u_i,v_i) = \left[ \begin{array}{ccc} e^{\mathrm{j}k({x_0}{u_i} + y_0{v_i})} & \ldots & e^{\mathrm{j}k({x_{N-1}}{u_i} + y_{N-1}{u_i})}  \end{array} \right]^{T}.
\end{equation}
The vector ${\mathbf{s}}$ contains the (unknown) complex signal source amplitudes and the vector ${\mathbf{b}}$ represents the complex output of the array or synthetic aperture.  The probe measures intensity or ${|[\mathbf{b}]_{k}|^2}$ at the ${k}$th spatial sample in the aperture.  The ${N}$ rows of ${\mathbf{A}}$ are electrical angle vectors that specify all possible phases at each array element over all angles in the search grid,
 \begin{equation}
 \mathbf{A} = \left[ \begin{array}{ccc} \mathbf{f}_1^T & \ldots & \mathbf{f}_{N-1}^T \end{array} \right]^{T}.
 \end{equation}
 For example,
 \begin{equation}
     \mathbf{f}_i = \left[ \begin{array}{ccc} e^{\mathrm{j}k({x_i}{u_0} + y_i{v_0})} & \ldots & e^{\mathrm{j}k({x_i}{u_{K-1}} + y_i{v_{K-1}})}  \end{array} \right]^{T}.
 \end{equation}
Since the probe measures electric field intensity at each array element,
 \begin{equation}
     |[\mathbf{b}]_{k}|^2 \equiv b_k^2 = \big|<\mathbf{f}_{k},\mathbf{s}>\big|^2 = \big|\mathbf{f}_{k}^{H}\mathbf{s}\big|^2.
 \end{equation}
The objective function to optimize is the maximum error between the intensity measurements and the magnitude squared signal model,
\begin{equation}
     \underset{{}}{\textrm{minimize}} \max_{k} \big|\thinspace \big|[\mathbf{b}]_{k}\big|^2 - \big|[\mathbf{As}]_{k}\big|^2\thinspace \big|
 \end{equation}
 or equivalently,
 \begin{equation}
     \underset{{}}{\textrm{minimize}} \max_{k} \big| b_k^2 - \big|\mathbf{f}_{k}^{H}\mathbf{s}\big|^2 \big|
 \end{equation}
for ${k = 0,\ldots,N-1}$.
This objective can be restated as
\begin{align}
    \textrm{minimize} \quad &\delta \\ \nonumber
    \textrm{subject to} \quad &\big|\thinspace |\mathbf{f}_{0}^{H}\mathbf{s}|^2 - b_{0}^2 \thinspace \big| \leq \delta  \\ \nonumber
    %&\big|\thinspace |\mathbf{f}_{1}^{H}\mathbf{s}|^2 - b_{1}^2 \thinspace \big| \leq \delta  \\ \nonumber
    &\quad \vdots  \\ \nonumber
    &\big|\thinspace |\mathbf{f}_{N-1}^{H}\mathbf{s}|^2 - b_{N-1}^2 \thinspace \big| \leq \delta  \\ \nonumber
    &\delta \geq 0.
\end{align}
% Equivalently, we can write
% \begin{align}
%     \textrm{minimize} \quad &\delta \\ \nonumber
%     \textrm{subject to} \quad &-\delta \leq  |\mathbf{f}_{0}^{H}\mathbf{s}|^2 - b_{0}^2  \leq \delta  \\ \nonumber
%     %\quad &-\delta \leq |\mathbf{f}_{1}^{H}\mathbf{s}|^2 - b_{1}^2 \leq \delta  \\ \nonumber
%     &\quad \vdots  \\ \nonumber
%     \quad &-\delta \leq |\mathbf{f}_{N-1}^{H}\mathbf{s}|^2 - b_{N-1}^2 \leq \delta  \\ \nonumber
%     &\delta \geq 0.
% \end{align}
Rearranging and expanding terms yields,
% \begin{align}
%     &\textrm{minimize} \quad \delta \quad \textrm{subject to} \\ \nonumber
%     &|\mathbf{f}_{0}^{H}\mathbf{s}|^2 - \delta \leq b_{0}^2, \quad -|\mathbf{f}_{0}^{H}\mathbf{s}|^2 - \delta \leq -b_{0}^2, \\ \nonumber
%     &|\mathbf{f}_{1}^{H}\mathbf{s}|^2 - \delta \leq b_{1}^2, \quad -|\mathbf{f}_{1}^{H}\mathbf{s}|^2 - \delta \leq -b_{1}^2, \\ \nonumber
%     &\qquad \vdots  \\ \nonumber
%   &|\mathbf{f}_{N-1}^{H}\mathbf{s}|^2 - \delta \leq b_{N-1}^2, \quad -|\mathbf{f}_{N-1}^{H}\mathbf{s}|^2 - \delta \leq -b_{N-1}^2, \\ \nonumber
%     &\delta \geq 0.
% \end{align}
% Continuing to expand terms results in,
\begin{align}
    &\textrm{minimize} \quad \delta \\   \nonumber
    &\textrm{subject to} \quad
    \mathbf{s}^H\mathbf{f}_{0}\mathbf{f}_{0}^{H}\mathbf{s} - \delta \leq b_{0}^2, \quad -\mathbf{s}^H\mathbf{f}_{0}\mathbf{f}_{0}^{H}\mathbf{s} - \delta \leq -b_{0}^2, \\ \nonumber
    %&\mathbf{s}^H\mathbf{f}_{1}\mathbf{f}_{1}^{H}\mathbf{s} - \delta \leq b_{1}^2, \quad -\mathbf{s}^H\mathbf{f}_{1}\mathbf{f}_{1}^{H}\mathbf{s} - \delta \leq -b_{1}^2, \\ \nonumber
    &\qquad \vdots  \\ \nonumber
   &\mathbf{s}^H\mathbf{f}_{N-1}\mathbf{f}_{N-1}^{H}\mathbf{s} - \delta \leq b_{N-1}^2, \thinspace -\mathbf{s}^H\mathbf{f}_{N-1}\mathbf{f}_{N-1}^{H}\mathbf{s} - \delta \leq -b_{N-1}^2, \\ \nonumber
    &\delta \geq 0.
\end{align}
Using the matrix trace results in,
\begin{align}
    &\textrm{minimize} \quad \delta \quad \textrm{subject to} \\ \nonumber
    &\Tr{\big[\mathbf{f}_{0}\mathbf{f}_{0}^{H}\mathbf{s}\mathbf{s}^H\big]} - \delta \leq b_{0}^2, \quad -\Tr{\big[\mathbf{f}_{0}\mathbf{f}_{0}^{H}\mathbf{s}\mathbf{s}^H\big]} - \delta \leq -b_{0}^2, \\ \nonumber
    %&\Tr{\big[\mathbf{f}_{1}\mathbf{f}_{1}^{H}\mathbf{s}\mathbf{s}^H\big]} - \delta \leq b_{1}^2, \quad -\Tr{\big[\mathbf{f}_{1}\mathbf{f}_{1}^{H}\mathbf{s}\mathbf{s}^H\big]} - \delta \leq -b_{1}^2, \\ \nonumber
    &\qquad \vdots  \\ \nonumber
   &\Tr{\big[\mathbf{f}_{N-1}\mathbf{f}_{N-1}^{H}\mathbf{s}\mathbf{s}^H\big]} - \delta \leq b_{N-1}^2, \\ 
   \nonumber
   &\hspace{0.3cm} -\Tr{\big[\mathbf{f}_{N-1}\mathbf{f}_{N-1}^{H}\mathbf{s}\mathbf{s}^H\big]} - \delta \leq -b_{N-1}^2, \\ \nonumber
    &\delta \geq 0.
\end{align}
To relax the constraints replace the rank-one matrix ${\mathbf{ss}^H}$ with the matrix ${\mathbf{S}}$ of general rank.  Applying this change and using the substitution ${\Tr{\big[\mathbf{A}^H\mathbf{B}\big]} = \vectz{\big(\mathbf{A}\big)^H}\vectz{\big(\mathbf{B}}\big)}$ yields,
\begin{align}
    \textrm{minimize} \quad &\delta  \\ \nonumber
    \textrm{subject to} \quad &\big[\widehat{\mathbf{A}}\widehat{\mathbf{s}}\big]_{k} - \delta \leq b_{k}^2 \quad k=0,\ldots,N-1 \\ \nonumber
    &-\big[\widehat{\mathbf{A}}\widehat{\mathbf{s}}\big]_{k} - \delta \leq -b_{k}^2 \quad k=0,\ldots,N-1 \\ \nonumber
    & \delta \geq 0 \nonumber
\end{align}
where ${\widehat{\mathbf{s}} = \vectz{\big(\mathbf{S}\big)}}$ and \begin{align}
    \widehat{\mathbf{A}} = \begin{bmatrix}  \vectz{\big(\mathbf{f}_{0}\mathbf{f}_{0}^H}\big)^T &
    %\vectz{\big(\mathbf{f}_{1}\mathbf{f}_{1}^H}\big)^T
    & \ldots & \vectz{\big(\mathbf{f}_{N-1}\mathbf{f}_{N-1}^H}\big)^T  \end{bmatrix}^T.
\end{align}

We can write this objective in the form of a standard linear program as
\begin{align}
    \textrm{minimize} \quad  &\left[ \begin{array}{cccc} 0 & 0 & \dots & 1 \end{array} \right] \left[ \begin{array}{c} \widehat{\mathbf{s}} \\ \delta \end{array} \right]  \\ \nonumber
    \textrm{subject to} \quad & \left[ \begin{array}{cc} \widehat{\mathbf{A}} & -1 \\ -\widehat{\mathbf{A}} & -1 \end{array} \right] \left[ \begin{array}{c} \widehat{\mathbf{s}} \\ \delta \end{array} \right] \leq \left[ \begin{array}{c} \mathbf{b}^2 \\ -\mathbf{b}^2 \end{array} \right] \\ \nonumber
    &\delta \geq 0.  \nonumber
\end{align}
Define the matrices
\begin{align}
    &\overline{\mathbf{A}} = \left[ \begin{array}{cc} \Re{\big(\widehat{\mathbf{A}}\big)} & -\Im{\big(\widehat{\mathbf{A}}\big)} \\ 
    \Im{\big(\widehat{\mathbf{A}}\big)} & \Re{\big(\widehat{\mathbf{A}}\big)} \end{array} \right], \\ \nonumber
    &\overline{\mathbf{b}} = \left[ \begin{array}{c} \mathbf{b}^2 \\ \mathbf{0} \end{array} \right], \quad
    \overline{\mathbf{s}} = \left[ \begin{array}{c} \Re{\big(\widehat{\mathbf{s}}\big)} \\ \nonumber \Im{\big(\widehat{\mathbf{s}}\big)} \end{array} \right].
\end{align}
Then the final optimization program takes the form,
\begin{align}
    \textrm{minimize} \quad  &\left[ \begin{array}{cccc} 0 & 0 & \dots & 1 \end{array} \right] \left[ \begin{array}{c} \overline{\mathbf{s}} \\ \delta \end{array} \right]  \\  \nonumber
    \textrm{subject to} \quad & \left[ \begin{array}{cc} \overline{\mathbf{A}} & -1 \\ -\overline{\mathbf{A}} & -1 \end{array} \right] \left[ \begin{array}{c} \overline{\mathbf{s}} \\ \delta \end{array} \right] \leq \left[ \begin{array}{c} \overline{\mathbf{b}} \\ -\overline{\mathbf{b}} \end{array} \right] \\ \nonumber
    &\delta \geq 0.  \nonumber
\end{align}
A regularization term to encourage a sparse solution can be added to the objective function as in
\begin{align}
    \label{E:eqn18}
    \textrm{minimize} \quad  &\left[ \begin{array}{cccc} 0 & 0 & \dots & 1 \end{array} \right] \left[ \begin{array}{c} \overline{\mathbf{s}} \\ \delta \end{array} \right] + \left\lVert \left[ \begin{array}{c} \overline{\mathbf{s}} \\ \delta \end{array} \right] \right\rVert_{1}  \\ \nonumber
    \textrm{subject to} \quad & \left[ \begin{array}{cc} \overline{\mathbf{A}} & -1 \\ -\overline{\mathbf{A}} & -1 \end{array} \right] \left[ \begin{array}{c} \overline{\mathbf{s}} \\ \delta \end{array} \right] \leq \left[ \begin{array}{c} \overline{\mathbf{b}} \\ -\overline{\mathbf{b}} \end{array} \right] \\ \nonumber
    &\delta \geq 0.  \nonumber
\end{align}

This program can be solved using interior point methods and optimization toolboxes such as CVX.  Once the complex matrix ${\mathbf{S}}$ has been reconstructed from the solution ${\overline{\mathbf{s}}}$, it is still necessary to determine the best rank-one approximation ${\mathbf{s}_{opt}}$ to ${\mathbf{S}}$.  One can refer to \cite{horn1999} for a proof that
\begin{equation}
    \label{E:eqn19}
    \mathbf{s}_{opt} = (\mathbf{v}^H\mathbf{S}\mathbf{v})\mathbf{vv}^H
\end{equation}
where ${\mathbf{v}}$ is the eigenvector of ${\mathbf{S}}$ corresponding to the largest eigenvalue.  Next, the final vector of complex signal source coefficients is used to determine the corresponding array output vector ${\mathbf{b}_{est} = \mathbf{A}\mathbf{s}_{opt}}$.  In the second and third stages of the algorithm described in Section  \ref{sec:algorithm}, alternating projections between the sets of feasible signal sources and array output magnitudes are used to search for ${e^{-\textrm{j}\angle{\mathbf{b}}}}$.
\section{Alternating Projections}
\label{sec:algorithm}
The second stage of the alternating projections algorithm is a projection into the set of feasible signal sources.  To implement this projection it is necessary to estimate or know the number of signal sources ${\widehat{K}}$ apriori.  Define the matrices,
\begin{align}
    \label{E:eqn21}
    &\widetilde{\mathbf{A}} = \left[ \begin{array}{cc} \Re{\big({\mathbf{A}}\big)} & -\Im{\big({\mathbf{A}}\big)} \\
    \Im{\big({\mathbf{A}}\big)} & \Re{\big({\mathbf{A}}\big)} \end{array} \right], \\  \nonumber
    &\widetilde{\mathbf{b}} = \left[ \begin{array}{c} \Re{\big(\mathbf{b}_{est}\big)} \\ \Im{\big(\mathbf{b}_{est}\big)} \end{array} \right], \quad
    \widetilde{\mathbf{s}} = \left[ \begin{array}{c} \Re{\big({\mathbf{s}}\big)} \\ \Im{\big({\mathbf{s}}\big)} \end{array} \right].
\end{align}
We can solve the lower-dimension linear program for ${\widetilde{\mathbf{s}}}$,
\begin{align}
    \label{E:eqn22}
    \textrm{minimize} \quad  &\left[ \begin{array}{cccc} 0 & 0 & \dots & 1 \end{array} \right] \left[ \begin{array}{c} \widetilde{\mathbf{s}} \\ \delta \end{array} \right]  \\  \nonumber
    \textrm{subject to} \quad & \left[ \begin{array}{cc} \widetilde{\mathbf{A}} & -1 \\ -\widetilde{\mathbf{A}} & -1 \end{array} \right] \left[ \begin{array}{c} \widetilde{\mathbf{s}} \\ \delta \end{array} \right] \leq \left[ \begin{array}{c} \widetilde{\mathbf{b}} \\ -\widetilde{\mathbf{b}} \end{array} \right], \quad \delta \geq 0.  \nonumber
\end{align}
The projection onto the set of feasible signal sources is implemented by retaining the ${\widehat{K}}$ largest coefficients ${\{\widetilde{s}_{0}, \widetilde{s}_{1}, \ldots, \widetilde{s}_{\widehat{K}-1}\}}$ of ${\widetilde{\mathbf{s}}}$ as in
\begin{equation}
    \label{E:eqn23}
    \mathbf{s}_{proj} = \left[ \begin{array}{ccccccc} \widetilde{s}_{0} & \widetilde{s}_{1} & \ldots & \widetilde{s}_{\widehat{K}-1} & 0 & \ldots & 0 \end{array} \right]^{T}.
\end{equation}
A new estimate of the array output vector is
\begin{equation}
\mathbf{b}_{1} = \left|\mathbf{b}\right| \odot e^{-\textrm{j}\angle\mathbf{A}\mathbf{s}_{proj}}.
\end{equation}
%\clearpage
\noindent\begin{minipage}{0.99\columnwidth}
\begin{algorithm}[H]
\begin{algorithmic}[1]
\vspace{4pt}
\INPUTS
\State{Construct a dense uniformly sampled grid of ${L}$ angles; ${\{(u_{l},v_{l}) \text{ } | \text{ } l = 0,\ldots,L-1\}}$.  Use this grid to create the array manifold matrix ${\mathbf{A}}$ given in (\ref{E:eqn02}).}
\State{Solve the optimization program in (\ref{E:eqn18}) for ${\overline{\mathbf{s}}}$ and rearrange the result to obtain ${\mathbf{S}}$.  Use (\ref{E:eqn19}) to obtain ${\mathbf{s}_{opt}}$ from ${\mathbf{S}}$.}
\State{Compute the corresponding array output vector ${\mathbf{b}_{est} = \mathbf{A}\mathbf{s}_{opt}}$.}
\State{Set the initial complex array output estimate to ${\mathbf{b}_{0}^{\dagger} = |\mathbf{b}| \odot e^{-\textrm{j}\angle\mathbf{b}_{est}}}$.  This step effectively projects the vector ${\mathbf{b}_{est}}$ onto the set of feasible array output magnitudes.}
\vspace{4pt}
\STAGEA
\end{algorithmic}\begin{algorithmic}[1]
\State{Use ${\mathbf{b}_{est}}$ to create the vector ${\widetilde{\mathbf{b}}}$ in (\ref{E:eqn21}).}
\State{Solve the linear program in (\ref{E:eqn22}) for ${\widetilde{\mathbf{s}}}$ and retain the ${\widehat{K}}$-largest coefficients as in (\ref{E:eqn23}) to create the vector ${\mathbf{s}_{proj}}$.}
\State{Update the complex array output estimate to ${\mathbf{b}_{1}^{\dagger} = \mathbf{b}_{0}^{\dagger} \odot e^{-\textrm{j}\angle\mathbf{A}\mathbf{s}_{proj}}.}$}
\vspace{4pt}
\STAGEB
\end{algorithmic}\begin{algorithmic}[1]
\State{Create the matrix ${\mathbf{R} = \mathbf{b}_{1}^{\dagger}\mathbf{b}_{1}^{\dagger{H}} + \nu\mathbf{I}}$ for small ${\nu > 0}$.}
\State{Initialize the conjugate gradient algorithm with ${\mathbf{w}_{0} = e^{-\mathrm{j}\angle\mathbf{A}\mathbf{s}_{proj}}.}$}
\State{Set ${\mathbf{g}_{0} = \mathbf{h}_{0} = \nabla{P}(\mathbf{w}_{0}).}$}
\State{For ${k=0,1,\ldots,J-1}$ iterations compute ${t_{k}}$ such that ${P\left(e^{-\mathrm{j}t_{k}\text{Diag}[\mathbf{h}_{k}]}\mathbf{w}_{k}\right) \geq P\left(e^{-\mathrm{j}t\text{Diag}[\mathbf{h}_{k}]}\mathbf{w}_{k}\right) \forall t \geq 0}$.}
\State{Set ${\mathbf{w}_{k+1} = e^{-\mathrm{j}t_{k}\text{Diag}[\mathbf{h}_{k}]}\mathbf{w}_{k}}$}.
\State{Set ${\mathbf{g}_{k+1} = \nabla{P}(\mathbf{w}_{k+1})}$.}
\State{Set ${\mathbf{h}_{k+1} = \mathbf{g}_{k+1} + \gamma_{k}\mathbf{h}_{k}, \quad \gamma_{k} = \frac{\left(\mathbf{g}_{k+1} - \mathbf{g}_{k}\right)^{T}\mathbf{g}_{k+1}}{\lVert\mathbf{g}_{k}\rVert^{2}}.}$}
\State{After ${N_{cg}}$ gradient iterations set ${\mathbf{b}_{est} = \mathbf{b}_{0}^{\dagger} = |\mathbf{b}|{\odot}e^{-\mathrm{j}\angle\mathbf{w}_{cg}}}$ and return to Step 1 in Stage 2 for the next linear program iteration.}
\State{After executing Stage 2 and Stage 3 for ${N_{ap}}$ iterations the final solution is ${|\mathbf{b}|{\odot}e^{-\mathrm{j}\angle\mathbf{w}_{ap}}}$}.
\end{algorithmic}
\caption{Three-Stage Phase Retrieval Algorithm}\label{ortho}
\end{algorithm}
\end{minipage}
\vspace{4pt}

We now seek to maximize the array output power,
\begin{equation}
\label{E:eqn25}
    P(\mathbf{w}) = |\mathbf{w}^{H}\mathbf{b}_{1}|^2 = \mathbf{w}^{H}\mathbf{b_{1}b_{1}}^{H}\mathbf{w} \equiv \mathbf{w}^{H}\mathbf{Rw}
\end{equation}
where ${\mathbf{w}}$ is a vector of phase weights,
\begin{equation}
    \mathbf{w} = \left[ \begin{array}{cccc} e^{j\phi_{0}} & e^{j\phi_{1}} & \ldots & e^{j\phi{N-1}} \end{array} \right]^T.
\end{equation}
The gradient vector ${\nabla{P}(\mathbf{w})}$ can be computed using an approach described in \cite{stevesmith1999} as,
\begin{equation}
\label{E:eqn27}
    \nabla{P}(\mathbf{w}) = \frac{1}{\mathbf{w}^{H}\mathbf{w}}\Im \big( \text{diag} \big[ \mathbf{R} - P_{s}(\mathbf{w})\mathbf{I}, \mathbf{ww}^H \big] \big),
\end{equation}
where ${[\mathbf{A}, \mathbf{B}] = \mathbf{AB} - \mathbf{BA}}$.  Conjugate gradient iterations can now be used to find an optimal solution for ${\mathbf{w}}$.

In the algorithm, notice that the input to Stage 2 from Stage 3, ${|\mathbf{b}|{\odot}e^{-\mathrm{j}\angle\mathbf{w}_{cg}}}$, is a projection onto the set of feasible array output magnitudes.  Likewise, the pruning operation in Step 2 of Stage 2 to create ${\mathbf{s}_{proj}}$ is a projection onto the set of feasible signal sources.  Thus the algorithm alternates projections between two constraint sets.  The primary purpose of Stage 1 is to provide a reliable and consistent starting point for the alternating projections.
\begin{figure}
 \begin{minipage}[b]{1.0\linewidth}
  \centering
  \includegraphics[width=6cm]{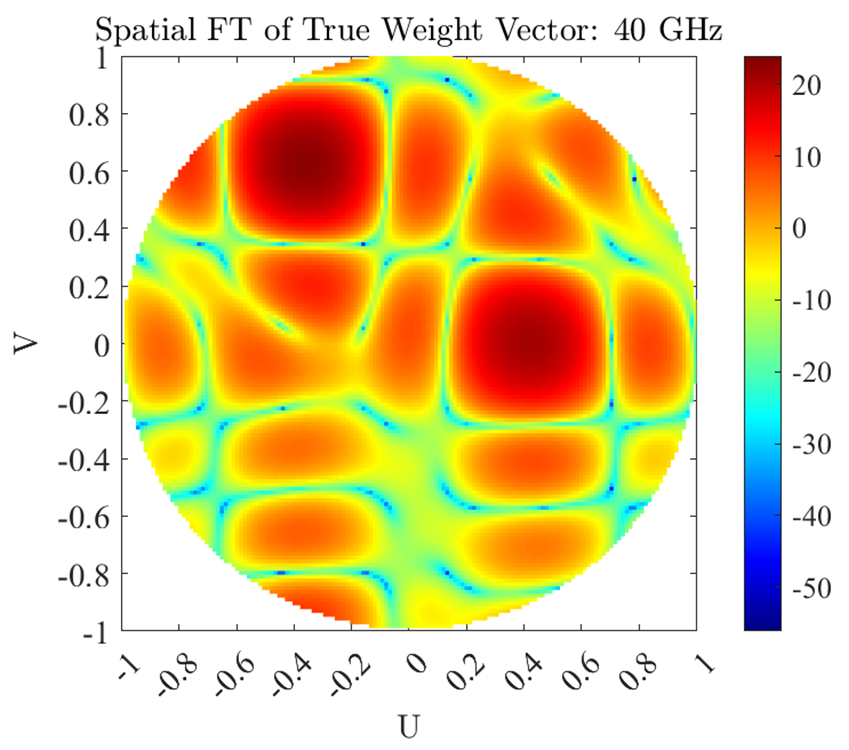}
  \vspace{0.2cm}
  \caption{Beamformed array output with known phase information}%\medskip
  \label{fig:truth}
% \end{minipage}
%\end{figure}
%\begin{figure}
% \begin{minipage}[b]{1.0\linewidth}
  \centering
  \includegraphics[width=6cm]{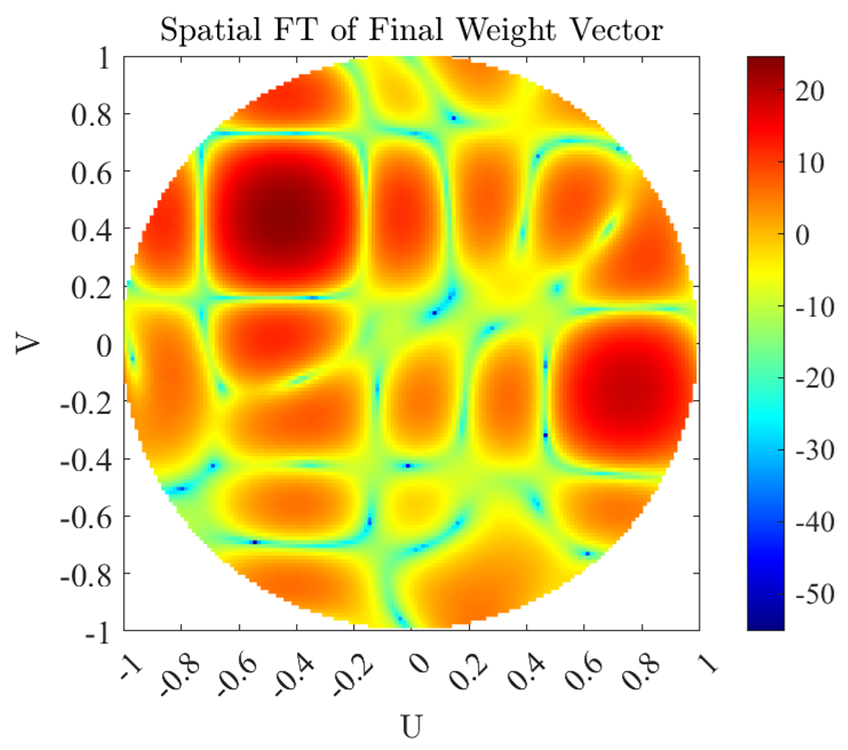}
  \vspace{0.2cm}
  \caption{Beamformed array output after estimating phase}%\medskip
  \label{fig:estimated}
% \end{minipage}
%\end{figure}
%\begin{figure}
% \begin{minipage}[b]{1.0\linewidth}
  \centering
  \includegraphics[width=6cm]{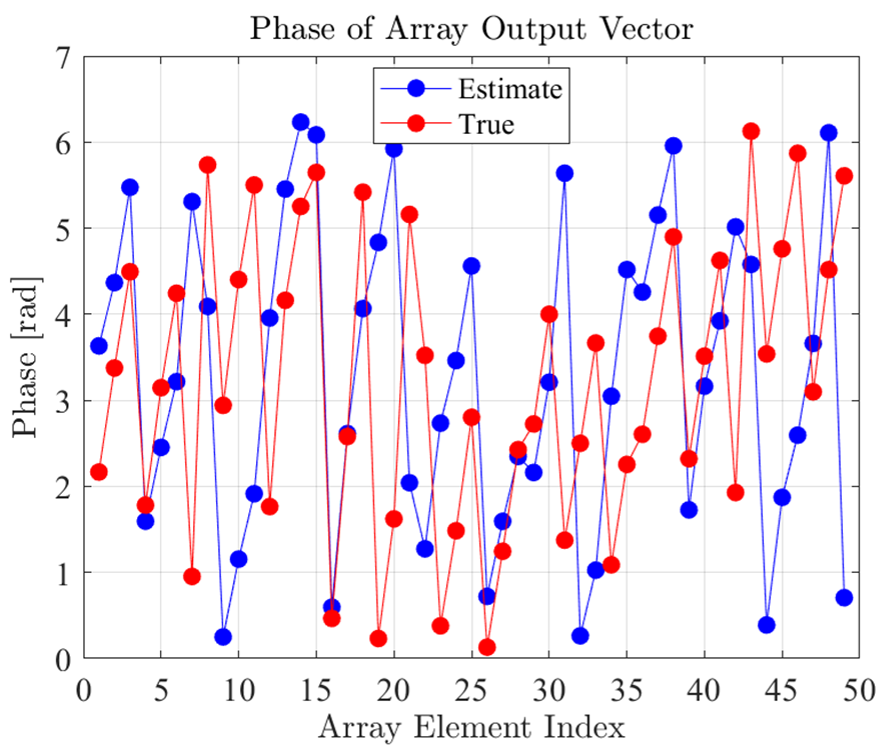}
  \vspace{0.2cm}
  \caption{True versus estimated phase across aperture}%\medskip
  \label{fig:phase}
% \end{minipage}
%\end{figure}
%\begin{figure}
% \begin{minipage}[b]{1.0\linewidth}
%  \centering
  \includegraphics[width=6cm]{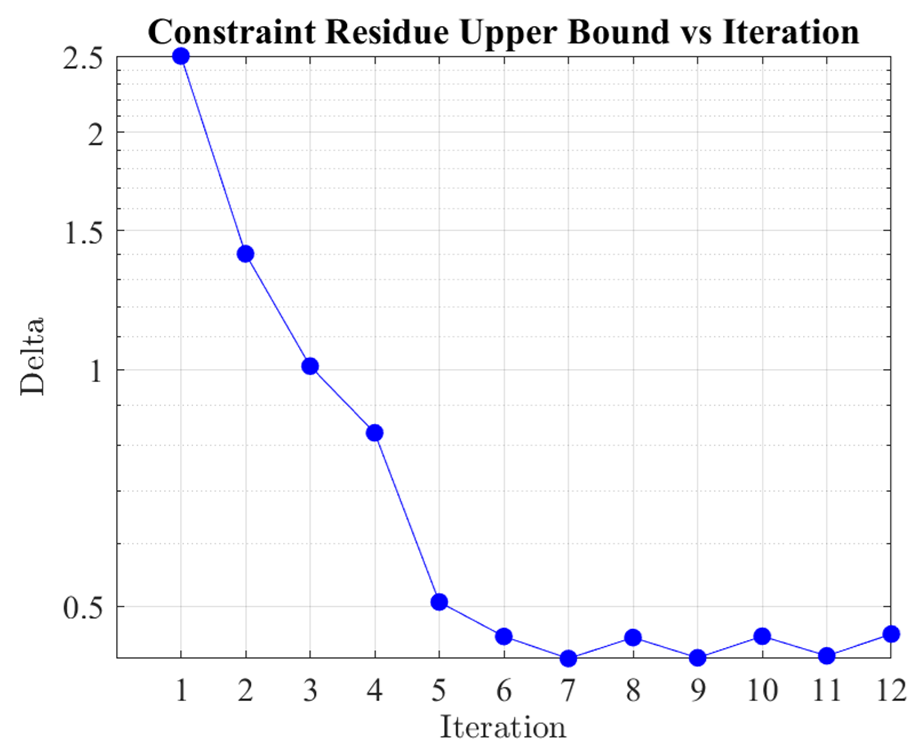}
  \vspace{0.2cm}
  \caption{Decreasing upper bound on maximum magnitude error (denoted as ${\delta}$ in the text)}%\medskip
  \label{fig:delta}
 \end{minipage}
\end{figure}
\section{Numerical Experiments}
\label{sec:numerical}
This section provides simulation results that demonstrate the phase retrieval algorithm.  Two signal sources of 10 and 8 dB are located at coordinates (-1663, 3155, 2870) and (1301, 2878, 0), in units of mm, with respect to a planar 7-by-7 synthetic aperture centered at the origin.  The spacing between spatial samples of the synthetic aperture is ${\lambda/2}$ at 40 GHz.  The total signal received at the ${m}$th spatial sample of the synthetic aperture is given by
\begin{equation}
    y_{m} = \sum_{k=0}^{\widehat{K}-1}\sqrt{P_{k}}e^{\textrm{j}(2\pi/{\lambda})d_{km}}
\end{equation}
where ${P_{k}}$ is the signal power and ${d_{km}}$ is the distance from the ${k}$th signal source to the ${m}$th spatial sample in the aperture.  In this scenario, the number of signal sources ${\widehat{K}=2}$.  Fig. \ref{fig:truth} illustrates the ground truth beamformed output created with perfect knowledge of the phase.  Fig. \ref{fig:estimated} illustrates the beamformed output when electric field intensity is measured and the phase is estimated.  The results show agreement between the estimated and true signal sources within one beamwidth.  Fig.\ref{fig:phase} illustrates the true and estimated unwrapped phase modulo ${2\pi}$ across the spatial samples of the synthetic aperture.  Fig. \ref{fig:delta} illustrates the progression of the scalar ${\delta}$ from (\ref{E:eqn22}).  %, which represents an upper bound on the error between the measurement model and the array output, for 12 iterations of the alternating projections algorithm.  The plot shows ${\delta}$ decreases steadily until a convergence plateau is reached.  Fig. 4 shows the value of the signal power cost function given in (\ref{E:eqn25}).  Over the course of 2e6 conjugate gradient iterations the cost function increases until a stationary point is reached where the gradient vector, given by (\ref{E:eqn27}), approaches zero.  Fig. 5 illustrates the norm of the gradient vector versus iterations of the conjugate gradient algorithm.

%In a second simulation scenario the 10 and 8 dB signal sources are at (-1663, 3155, -5500) and (1301, 2878, 4000) respectively.  Fig. 7 depicts the ground truth beamformed output and Fig. 8 shows the results using the retrieved phase across the aperture.  Again in this scenario the estimated signal sources are within one beamwidth of the true locations but in this case the raw output of the phase retrieval algorithm corresponded to an array rotated by 90 degrees.  Fig. 9 plots the unwrapped estimated and true phase across the synthetic aperture modulo ${2\pi}$.

\section{Summary}
\label{sec:summary}
%In this paper a novel phase retrieval algorithm is derived for use in synthetic apertures that measure electric field intensity using Rydberg quantum probes.
In this paper a novel alternating projections algorithm is derived for estimating signal phase at each spatial sample of a synthetic aperture where the receive probe is a Rydberg quantum sensor that measures electric field intensity. % The distance and power of the signal sources is unknown but the number of signal sources is assumed known.  Simulation results for the noiseless case show that the algorithm estimates the signal sources to within a beamwidth of their true locations for a multitude of geometries.
%\clearpage
% \begin{figure}
% % \begin{minipage}[b]{1.0\linewidth}
%   \centering
%   \includegraphics[width=5.5cm]{cost_vs_iter.png}
%   \vspace{0.5cm}
%   \caption{Objective function versus conjugate gradient iteration}%\medskip
%   \label{fig:sparse_array4}
% % \end{minipage}
% \end{figure}
% \begin{figure}
% % \begin{minipage}[b]{1.0\linewidth}
%   \centering
%   \includegraphics[width=5.5cm]{grad_vs_iter.png}
%   \vspace{0.5cm}
%   \caption{Norm of gradient vector versus conjugate gradient iteration}%\medskip
%   \label{fig:sparse_array5}
% % \end{minipage}
% \end{figure}
% \begin{figure}[htb]
% % \begin{minipage}[b]{1.0\linewidth}
%   \centering
%   \includegraphics[width=7.5cm]{truth_Case_2.png}
%   %\vspace{0.5cm}
%   \caption{Ground truth beamformed output -- Case 2}%\medskip
%   \label{fig:truth_Case_2}
% % \end{minipage}
% \end{figure}
% \begin{figure}[htb]
% % \begin{minipage}[b]{1.0\linewidth}
%   \centering
%   \includegraphics[width=7.5cm]{reorient_Case_2.png}
%   %\vspace{0.5cm}
%   \caption{Beamformed output using retrieved phase -- Case 2}%\medskip
%   \label{fig:reorient_Case_2}
% % \end{minipage}
% \end{figure}
% \begin{figure}[htb]
% % \begin{minipage}[b]{1.0\linewidth}
%   \centering
%   \includegraphics[width=7.5cm]{phase_across_aperture_Case_2.png}
%   %\vspace{0.5cm}
%   \caption{Estimated versus true phase across aperture -- Case 2}%\medskip
%   \label{fig:phase_Case_2}
% % \end{minipage}
% \end{figure}

%\clearpage
\bibliographystyle{IEEEtran}
\bibliography{references,refs2}

\end{document}